# Behavior of implanted Xe, Kr and Ar in nanodiamond and thin graphene stacks: experiment and modeling

Andrey A. Shiryaev,[*,a] Alexander L. Trigub,[b] Ekaterina N. Voronina,[c]

Kristina O. Kvashnina,[d,e,f] Valentin L. Bukhovets[a]

*a. A.N. Frumkin Institute of Physical Chemistry and Electrochemistry RAS, Leninsky pr. 31 korp. 4, 119071, Moscow, Russia. E-mail: shiryaev@phyche.ac.ru, a_shiryaev@mail.ru*
*b. National Research Center «Kurchatov Institute», Moscow, Russia*
*c. Department of Physics Lomonosov Moscow State University, 119991 Moscow, Russia*
*d. The Rossendorf Beamline at ESRF – The European Synchrotron, CS40220, 38043 Grenoble Cedex 9, France*
*e. Helmholtz Zentrum Dresden-Rossendorf (HZDR), Institute of Resource Ecology, PO Box 510119, 01314 Dresden, Germany*
*f. Department of Chemistry, Lomonosov Moscow State University, 119991 Moscow, Russia*

**ABSTRACT.** Implantation and subsequent behaviour of heavy noble gases (Ar, Kr, Xe) in few-layer graphene sheets and in nanodiamonds is studied both using computational methods and experimentally using X-ray absorption spectroscopy. X-ray absorption spectroscopy provides substantial support for the Xe-vacancy (Xe-V) defect as a main site for Xe in nanodiamond. It is shown that noble gases in thin graphene stacks distort the layers, forming bulges. The energy of an ion placed in between flat graphene sheets is notably lower than in domains with high curvature. However, if the ion is trapped in the curved domain, considerable additional energy is required to displace it.

## Introduction

Nanoparticles of sp2- and sp3-hybridised carbon are frequently encountered both in nature and in various technologies. Many types of poorly ordered $sp^2$ carbons comprise stacks of few graphene sheets of different sizes. Interestingly, natural nanocarbons encountered in some meteorites contain relatively high abundances of noble gases (review in Ref. 1). Ion implantation is the most feasible mechanism of incorporation of these atoms into nanodiamonds. A peculiar and rather abundant component of noble gases in meteorites - the Q-component ("Q" stands for "quintessence") is likely a disordered carbonaceous phase with abundant few-layers graphene stacks.[2-6] It is suggested that noble gases reside between the sheets and although the mechanism(s) of the gases' incorporation is not firmly established, their isotopic ratios strongly suggest that low energy ion implantation of ionized species was very important. One of the most unusual features of the Q-phase is rather strong binding of trapped noble gases (He, Ar, Xe): during step oxidation these gases are released only at 450-500 °C, i.e. shortly before burning of the carbon carrier.[6] Since atomic radii of heavy noble gases (Kr, Xe) exceeds interlayer spacing in graphene stacks present in $sp^2$-C from meteorites, the graphene sheets deform, keeping the atom in the bulge. It was suggested that high retention capacity is explained by peculiarities of the atomic structure of sites where trapped noble atoms reside.

Ion implantation into nanostructures attracts considerable attention.[7] Numerous studies address the interaction of ions with single layer graphene[8-13], but these works primarily focus on radiation defects in single layer graphene or the formation of nanopores, leaving aside thin graphene stacks. Xe implantation in nanodiamond grains was studied using in situ TEM and different computational approaches.[14-16] However, direct information about the local environment of noble ions implanted into nanodiamonds and various $sp^2$-carbons is very limited.

In this contribution, we present results of experimental and computational investigation of the local environment of Xe, Kr and Ar ions implanted with energies below 1500 eV into several types of nanocarbons. Light gases (H and He), and graphene stacks decorated by functional groups are briefly considered for comparative purposes. The paper is organized as following: at first, we discuss computational modeling of the implantation process and relaxed structures of the noble gas atoms in flat and curved graphene stacks obtained by quantum chemistry and then compare theoretical modeling with experimental data on the local environment of implanted ions.

## Sample and methods

*Computational methodology*

Molecular dynamics (MD) simulations of the noble gas ion interaction with flat and curved graphene sheets were carried out with LAMMPS codes.[17]. To simulate ion implantation process AIREBO bond-order force field, which enables to take into account formation and rupture of chemical bonds and includes additional terms to take into account van der Waals interaction between graphene sheets,[18] was used to describe C–C interaction. The C–Ar and C–Xe interactions were described by the ZBL potential.[19] To study the behaviour of implanted ions within graphene stacks we applied ReaxFF force field developed especially for C–(He, Ne, Ar, Kr) interactions.[20]

The initial model of a graphene sheet was 5.03×4.84 nm$^2$ in size and consisted of 960 carbon atoms. On this base, two main models of multi-layered graphene (2 and 4 sheets) were built. A few carbon atoms in the corner of each sheet model were fixed, and periodic boundary conditions were applied in XY directions. Importantly, the sample was free-standing, i.e. no substrate was employed.

Implantation process was simulated as following. At the initial moment of the time cycle, an ion was created with a given kinetic energy above the surface. Initial ion (X, Y) coordinates were chosen randomly but with a condition that the distance of the ion from the central axis of the model does not exceed 0.5 nm. Ions were directed along the Z axis towards the model at normal incidence. The energy of ions was varied from 20 to 200 eV with a step of 5 eV, and for each energy value 100 impacts were simulated. To keep the overall temperature of the system at 50 K, the Berendsen thermostat was used periodically in accordance with the algorithm proposed in Ref. 21: for the first 1.0 ps the simulation was performed in the microcanonical (NVE) ensemble with no temperature control to guarantee a proper development of heat transfer and defect formation, then the thermostat was switched on in order to cool the model down to 50 K. In the first part of this cycle, the time step was chosen equal to 0.1 fs, and in the second part with the thermostat applied, it was increased up to 0.5 fs, so the total duration of one cycle was 10 ps.

To simulate the behaviour of implanted ions within curved graphene stacks, two additional models were built. They are based on 2- or 4-layered graphene sheets joined with open semi-nanotubes, i.e. a nanotube was cut in two halves by a plane parallel to the main axis. To preserve both flat and curved zones of such models during the relaxation, some atoms in each layer were fixed. Dynamic simulations were carried out in NVT ensemble at 100, 200 and 300 K; no additional kinetic energy was supplied to the implanted ion. The MD simulations were performed on the equipment of the shared research facilities of High Performance Computing resources at Lomonosov Moscow State University.[22] For the visualization of models and analysis of simulation results, the software package OVITO was used.[23]

The behaviour of Xe and Kr in nanodiamonds was modelled with non-boundary conditions in cubic unit cells, which provided distances between diamond particles more than 10 Å to avoid interparticle interaction. Nanoparticles were constructed as described in Ref. 24 and allowed to relax. Subsequently, a grain with a Xe-V defect was optimized. The Xe-V defects were built by replacing two neighbouring carbon atoms (see also Ref. 14). For evaluation of local atomic geometry of Xe(Kr) in $sp^2$ carbons we performed DFT geometry optimization of Xe(Kr) in two and four sheets of graphene using the plane-wave-pseudopotential approach as implemented in Quantum ESPRESSO.[25] The core electrons are described by the non-conserving Goedecker-Teter-Hutter (GTH) pseudopotentials.[26] In order to account for the van der Waals interactions between graphene sheets the revised Vydrov-Van Voorhis non-local correlation functional (rVV10)[27] was applied. The convergence criterion for self-consistent energy is taken to be $10^{-6}$ Ry. A Γ-point approximation was applied for the supercell of 6×6 graphene sheets. The kinetic energy cut-off for the electron wave functions is set at 100 Ry and the augmented charge density cut-off is set to 400 Ry. The ion of interest was introduced in between the layers and allowed to relax.

To interpret experimental XANES spectra quantum chemistry calculations were performed to provide initial structural models of Xe(Kr). The Quickstep module of the CP2K program suite with a dual basis of localized Gaussians and plane waves was employed.[28] The plane wave cutoff was 400 Ry, appropriate for employed GDH pseudopotentials.[26] The localized basis set of double-zeta plus polarization (DZVP) was quality optimized to reduce the basis set superposition errors.[29] The calculations were performed using the Perdew–Burke–Ernzerhof (PBE) exchange correlation functional.[30] A conjugation gradient (CG) geometry optimization with SCF convergence criteria of $5.0 \times 10^{-7}$ a.u. was used. Atomic configurations were considered converged when forces were less than $4.5 \times 10^{-4}$ hartree×bohr$^{-1}$. Theoretical XANES spectra were calculated with FDMNES code[31] using the modelled local environment of the Xe(Kr) atoms.

**Experimental**

Two sets of nanodiamonds with well-defined grain sizes (5 and 40 nm), nanocrystalline diamond film (UNCD, see Ref. 32 for detail), graphite, and technical soot were implanted using Kr/Xe asymmetric capacitive discharge at a dedicated setup equipped with an 2 MHz RF generator at IPCE RAS. The nanodiamond samples were either embedded into high purity oxygenless copper or placed into high purity graphite boats for Atomic Emission spectroscopy; Si and In substrate were also employed in some runs. Pressure of target gases (Xe or Kr) was between 3 and 6 Pa. At first, the samples were pre-sputtered under bias of -200-300 V for 10 min. Then the generator anode current was increased and the samples were implanted under bias of -1420-1490 V for Xe and ~-1000 V for Kr. We recall that considerable fraction of ions in the sheath possesses energies smaller than the maximum in eV, numerically equal to the magnitude of the bias in Volts.[33] The maximum energy of the ions is comparable with previous studies[34,35] and the penetration depth (1-3 nm) of the impinging ions was less than a grain diameter in all nanodiamond samples.

The selection of the ion fluence is a controversial task. For the XANES measurements, a relatively high concentration of the target ion (Xe/Kr) is desirable. At the same time, nanoparticles can be strongly heated or even completely destroyed by a single ion[14-16] and thus sputtering of upper layers of nanograins will occur. In our experiments the fluence was between $5-8\times10^{15}$ at/cm$^2$ with a dose rate of $\sim 8\times10^{12}$ at/cm$^2$·sec. The temperature of the samples was not controlled, but was certainly less than the indium melting temperature (156 °C). To preserve the implanted layer from dispersion, the samples were covered by an SPI® cellulose acetate replicating film, hardened by acetone immediately after the opening of the implantation vessel to the atmosphere.

The XANES measurements were performed at the ROBL beamline at ESRF.[36] Holders with implanted samples were placed on a dedicated Grazing-incidence setup. The measurements at the Xe-L3 and Kr-K-edges were performed in fluorescence mode by using 12-elements Ge detector. The estimated incident flux was $2\times10^{11}$ ph/sec at the Kr K-edge and $1.6\times10^9$ at the Xe L3 edge. Due to very low concentrations of implanted ions only near-edge spectra were recorded; up to 20 repeated scans were acquired for each sample. Data treatment was performed using Athena software.[37]

## Results

### *Modeling of Ar/Kr/Xe behavior in flat and curved graphene stacks*

At energies of the incident Xe ion below ~60 eV it is reflected by the graphene stacks, but leads to displacement of some carbon atoms. At ~60 eV the recoils may form links between the layers. A Xe ion penetrates the upper layer of a graphene stack at a relatively high energy of ~70 eV for bi-graphene and ~65 eV for 4-layer stack. The higher energy required for penetration into the bigraphene is likely explained by greater flexibility of thinner stack and, consequently, a larger degree of bending, which dissipates the energy of the incident ion. With increasing energy the penetration probability increases, e.g., at 80 eV 18(26)% of ions penetrate the bi- (4-)layer graphene, at 90 eV – 55%, at 100 eV – 90% and at energies exceeding 110 eV the upper layer is always penetrated. At yet higher energies, the ion also passes through the second graphene sheet thus leaving the bigraphene sample or is trapped in between layers of thicker stacks.

It is worth noting that energy dissipation occurs more slowly for bigraphene model in comparison with thicker stacks; such an effect is expected in 2D nanostructures because of their reduced dimensionality.[7] Figure 1 shows snapshots of bi- and 4-layer graphene stacks after a single impact of a 100 eV Ar ion. The ion implanted between the two uppermost layers induces distortion in the 3rd and 4th layers and the magnitude of the deformation decreases faster in the 4-layer stack.

Due to relatively high residual energy, an ion that has penetrated the upper graphene sheet starts to migrate rapidly in the gap between the layers. In the absolute majority of cases, the ion diffuses away from a vacancy-type defects formed during the implantation process. Therefore, calculation of the most stable configuration of trapped Xe is effectively reduced to a "static" case, which is discussed below together with experimental XANES data. Representative stable configurations of Xe(Kr) in flat bi- and 4-layer graphene stacks optimized using the DFT Quantum Espresso package are shown in Fig. 1. For Xe the closest neighbours are at 2.99-3.03 Å in bilayer and at 2.91-2.96 Å in four-layers stack. For Kr the values are 2.79-2.88 Å and 2.74-2.82 Å, respectively. Experimentally measured interlayer spacing in graphene stacks in carbon extracted from a meteorite is approx. 3.4-3.5 Å.[6] As shown by the simulation results, the emplacement of a large ion in between the sheets leads to the formation of a bulge. The heavier the ion is, the larger this bulge becomes, and, correspondingly, the energy change induced by the ion implantation especially in the case of heavy Xe and Kr ions increases. This fact indicates that the movement of an implanted ion in the stack is hindered, especially in in the case of heavy Xe and Kr ions.

**Fig. 1.** Dynamics of energy dissipation of 100 eV Ar ion impact in bi- and 4-layer graphene stacks

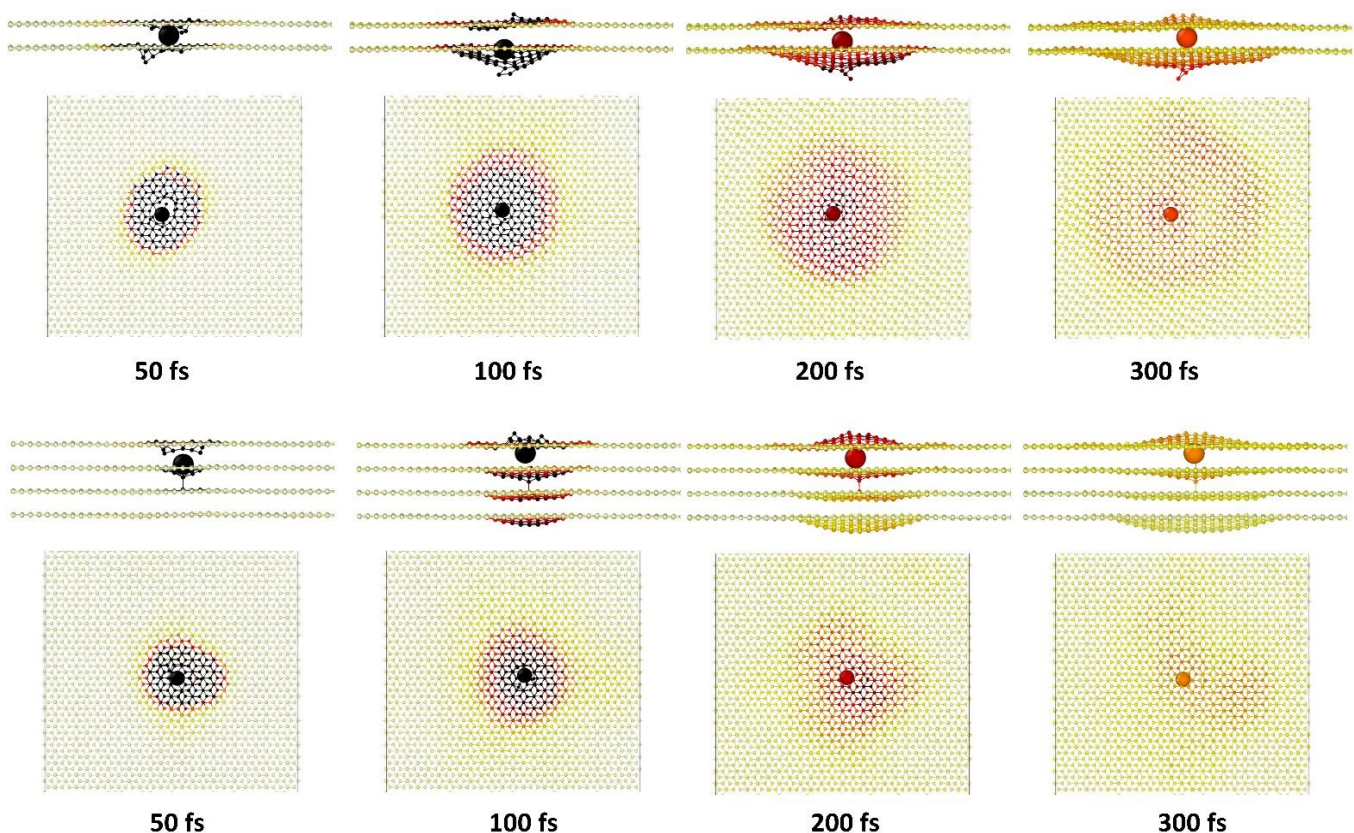

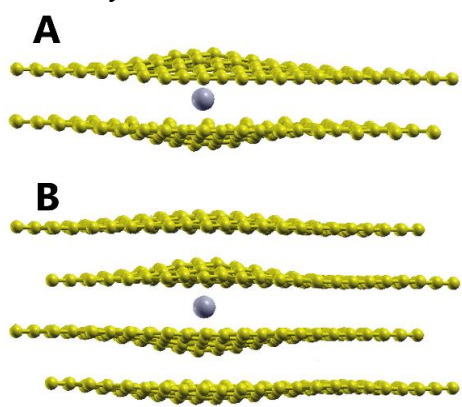

**Fig.2.** Relaxed stable configurations of Xe ion in bi- and 4-layer graphene.

Since the current work is partly driven by cosmochemical applications, it is important to consider curved graphene stacks, which are abundant in carbonaceous residues from meteorites.[2-6] Several possible models, based either on short two-wall nanotubes of different radii and lengths, or on nested nanotubes with a closed end and a graphene sheet, were studied. The simulations were run both with fixed and free edges for all types of models. The main effect of an implanted ion on the CNT structure was similar to that of the flat graphene stack: as a result of the implantation, the nanotube was significantly deformed, as shown in Fig. 3a, b. Because of different model structure and the number of atoms, direct comparison of energy change between planar and curved models is difficult. However, our DFT calculations show that for small models the implantation of noble gas ion into a curved structure leads to a higher excess energy than into a planar one (Table 1). This excess energy $\Delta E$ was calculated for the model of flat bilayer graphene and two-wall CNT of similar size as the difference between the DFT total energies of the particular model with an implanted atom and the model with the atom outside. Its positive values imply that the implantation is energetically unfavourable and requires a large amount of energy, so it can be considered as a measure of the graphene/CNT deformation. For example, for a two-walled CNT with the outer diameter of ~1 nm and the length of 1.5 nm the excess energy due to the Xe implantation amounts to ~7.3 eV, while for planar bilayer graphene of similar size it is equal to ~5.9 eV. Similar effect was found Taken together, the DFT modelling results indicate that curved graphene structures are energetically less favourable for the accommodation of large ions in comparison with the flat ones, see Table 1.

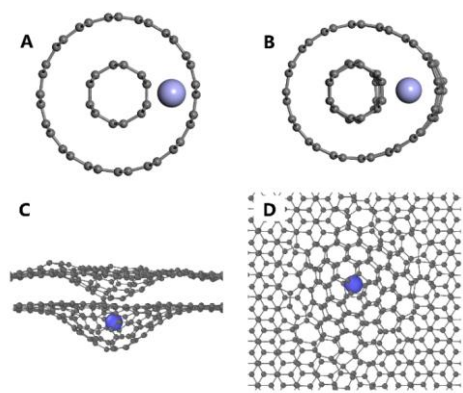

**Fig. 3.** Xe configurations in systems with high curvature. Two-walled periodic carbon nanotube (CNT): A – before, B - after the structure optimization. C-D: nested open CNT/graphene structure; side and top views of the optimised system, respectively.

**Table 1.** Energy increase ΔE due to the implantation of a single noble gas ion.

| Ion | ΔE (eV) | |
|---|---|---|
| | Flat bi-graphene | Two-wall CNT |
| Ar | 3.8 | 4.4 |
| Kr | 4.1 | 4.7 |
| Xe | 5.9 | 7.3 |

MD calculations carried out for the larger model showed that the stress induced by an implanted ion is redistributed over the whole model. Therefore, an additional, relatively large model of 4-layered graphene with planar and curved regions was built on the base of joined graphene stacks and CNTs with a diameter of 1.3 nm (Fig. 4). In order to prevent the model transformation during the relaxation, some carbon atoms on the edges were fixed. Then the motion of an implanted ion initially placed between layers in the centre of the flat and curved regions (Fig. 4 a and b, respectively) was studied at steady-state conditions with NVT ensemble.

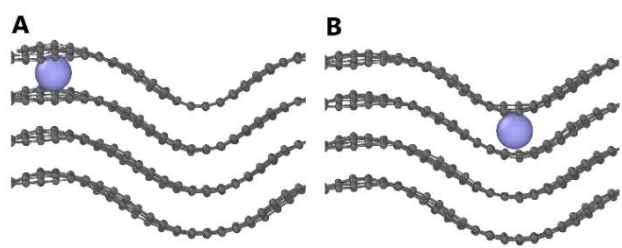

**Fig. 4.** An Ar ion in the planar (A) and curved (B) regions of the 4-layer graphene model after the relaxation (the x-axis directed to the right, see text).

MD simulations run at 300 K demonstrated that the ion placed in the planar region is able to move in the interlayer space, for instance, it can go to the leftmost part of the model (Fig. 4a). The situation changes drastically, when the ion is placed in the curved region (Fig. 4b). In this case, it just oscillates near the lowest point with a relatively small amplitude; the fate of the implant depends on the ion mass: relatively heavy Kr ions are effectively trapped in the dip, whereas lighter ions like Ar may still leave it. Fig. 5 demonstrates the dependencies of the x-coordinate of implanted ions with time during our simulations. As one can see, the Ar ion readily moves in the flat region, but not in the curved one (blue vs. purple lines in Fig. 5a), while the movement of the larger Kr ion is limited and demonstrates a similar pattern in both regions (Fig. 5b). In long simulation runs the Kr ion may, however, jump to neighbouring positions being in the flat region (see the right end of the blue line in Fig. 5b), but is still immobile in the curved one. One can explain this result by considering p-orbitals of carbon atoms: in planar graphene they are directed perpendicular to the sheet, whereas in curved graphene or CNTs these orbitals are inclined.[38] As a result, the electron density distribution in concave curved graphene regions becomes higher, and additional kinetic energy is required to overcome the barrier. The height of this barrier depends on the size and curvature of the model as well as the ion type. According to our simulations with the ReaxFF it is in order of few tenths of eV. Due to the same reason if the ion nevertheless manages to leave the curved region, it does not readily return back and remains in the flat part.

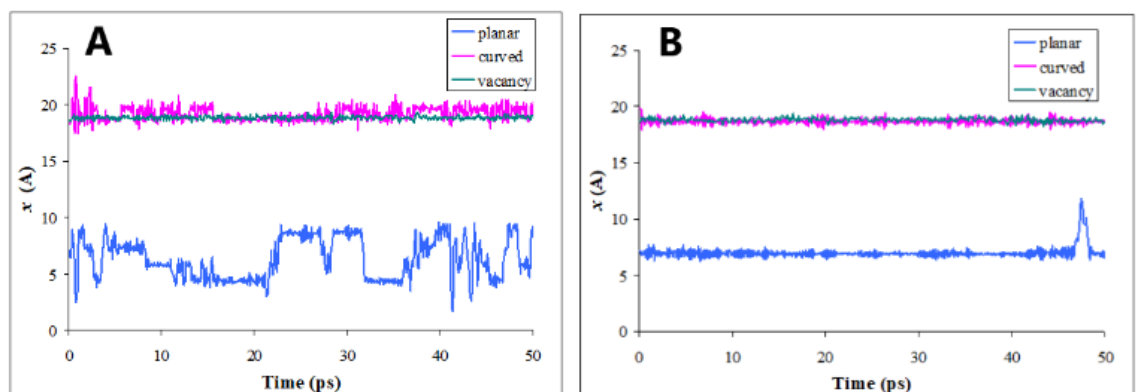

**Fig. 5.** Time dependencies of the x-coordinate of the implanted Ar (a) and Kr (b) ions implanted into of the 4-layered graphene model shown in Fig. 4. Blue and purple lines correspond to ions implanted into planar and curved regions of the model, respectively; green line corresponds to the implantation near a vacancy.

Interestingly, if we consider a noble gas ion in the vicinity of a (pre-existing) carbon vacancy, movements of this ion are quite limited and the ion prefers to stay near the defect. This behaviour is illustrated with green curves in Fig. 5 corresponding to the movement of ions implanted near the vacancy in the curved region. For the larger Kr ion the dependencies remain almost the same, while the behaviour of the lighter Ar ion change significantly (compare purple and green lines in Fig. 5a). Taken together with the results given above, we conclude that although the placement of a noble gas ion in between curved graphene sheets is energetically more expensive than the position in a flat region, if the ion is nevertheless getting into the curved region, it will likely stay there. The trapping effect is more pronounced when the curved sheets are defective.

Although the present study mainly focus on medium-to-heavy ions, for comparative purposes it is instructive to consider also hydrogen and helium. Impacts of low-energy He ions and H ions/radicals on graphene were simulated in some DFT[39] and MD[7,10,13,40] calculations. In general, due to smaller size and mass, helium ions are more prone to backscattering at lower energies and easier penetrates through graphene sheets is sufficient energy is provided. As a result, the probability to find an implanted He ion in the bi-layered graphene is substantially lower than for Ar ions (in the energy range when the penetration is possible). Moreover, implanted He atoms do not induce significant deformation of the graphene stack: the excess energy for He in flat be-graphene is ~1.2 eV, which is much less than for heavier noble gas atoms.

Although hydrogen is not a noble gas, it is important to discuss its behaviour, in particular, as a kind of a bridge from ideal "clean" graphene stacks to more realistic case of graphene decorated with functional groups. In contrast to chemically inert noble gas atoms, hydrogens are reactive and tend to form chemical bonds with carbon atoms.[40,41] When neutralized atom enters the interlayer space it interacts with neighboring C atoms forming a C–H bond and highly localised site for the implant. The subsequent change in the hybridization of the C atom from $sp^2$ to $sp^3$ induces local deformation of the graphene sheet near the adsorption site (Fig. 6).

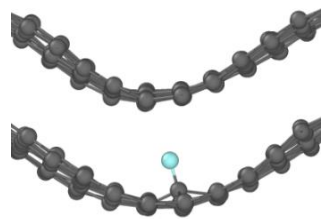

**Fig. 6**. Final position of a hydrogen ion implanted in bi-graphene.

Graphene stacks discussed above represent an important case of chemically pure monoelemental 2D material. However, graphene sheets in natural environments are at least partly decorated by various functional groups. Remarkable richness of possible functional groups drives comprehensive study of even the most important atomic configurations far beyond limits of the current study, but some hints of possible influence of the functionalization on behavior of implanted gases are presented below.

In case of flat and curved graphenes, many functional groups destroy the $sp^2$ hybridization of C atoms, resulting in local change of physical properties.[42] To analyze eventual influence of surface groups on our results, additional MD calculations with AIREBO force field were performed for the larger 4-layered graphene model. The upper graphene sheet was covered with H atoms (surface coverage is ~1%), resulting in formation of surface CH groups, in which C atoms become $sp^3$ hybridized. As a result of the relaxation, an implanted Ar ion stays close to the site with the C–H groups (Fig. 7). Functional groups on graphenes can be considered as defects and thus influence preferential positions of implanted ions. For the particular case of hydrogen studied here, the effect is smaller than in the case of vacancies. However, adatoms on graphene sheets (including cases with large curvature such and CNTs) show different behavior, for example, oxygen may be located above the C-C bond forming epoxy group or, at higher concentrations, several epoxy groups may convert to ethers, breaking underlying C-C bond.[43-46] In the latter case the sheet deformation is more pronounced and extensive and the implanted ion may migrate to the relevant site. In particular, in Ref. 6 it was shown that that upon $H_2O_2$ oxidation of meteoritic carbon, the noble gases are retained in the sequence Xe<Ar«He, while after $HNO_3$ the sequence changes to He«Xe<Ar. Large deformation of graphene sheets by adatoms and functional groups serves a plausible explanation of this behavior.

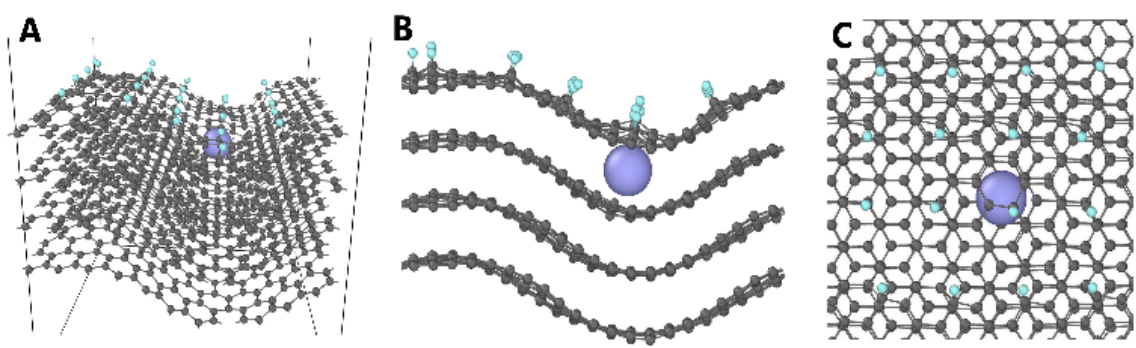

**Fig. 7.** An implanted Ar ion in a curved 4-layer graphene stack with the upper surface covered with C-H groups. The model is relaxed after the "implantation". A – general view, B – side view, C – top view.

## Experimental XANES spectra of Xe in nanocarbons

### Xe in nanodiamonds

Upon suitable annealing, xenon ions driven into bulk diamond by ion implantation may form stable defects responsible for several lines in photoluminescence spectra.[47,48] According to computational studies, the most stable atomic configuration of a large ion (X) in diamond lattice is an ion-vacancy complex, commonly denoted as X-V. According to several independent works the Xe-V defect is the most stable configuration for this impurity in diamond.[14,16,49,50] However, experimental confirmation of this model are yet lacking, since detailed investigations of Zeeman splitting and polarization dependence of photoluminescence of known Xe-related defects do not allow unambiguous selection of a defect model.[51,52]

Figure 8a shows a diamond lattice with a Xe-V complex as obtained by QuantumEspresso modelling.[14] The nearest C neighbours are at 2.17 Å (for a 2 nm grain), which is very close to 2.15 Å obtained with DFT for the defect in bulk diamond[44], but somewhat smaller than 2.25 Å obtained for the Xe-V in a nanodiamond grain with size of 3.9 nm.[16] The difference between values obtained for nanosized grains might be due to different assumptions of the surface structure of the grains: H-terminated surfaces were used in Ref. 14, whereas calculations in Ref. 16 were made for a cuboctahedron grain with different types of surface reconstruction of {100} and {111} faces.

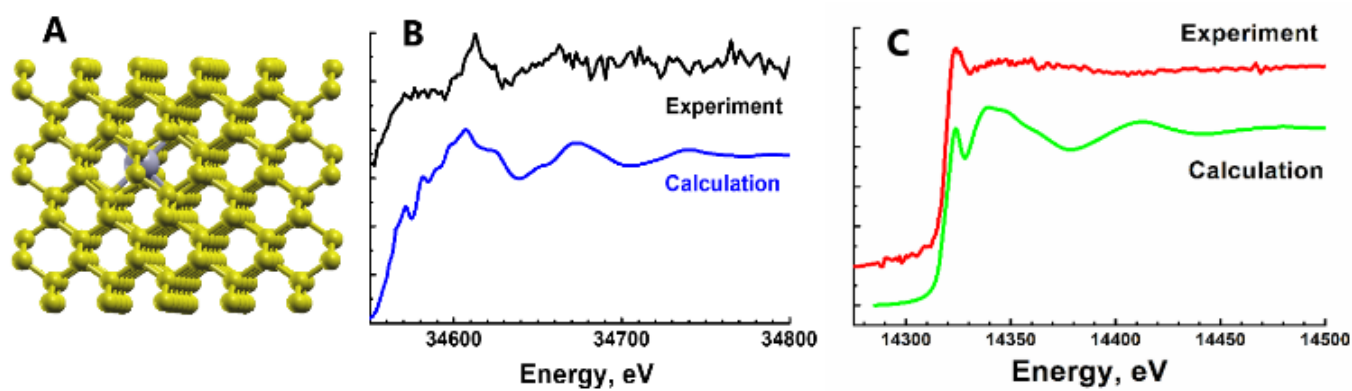

**Fig. 8.** Xe and Kr in nanodiamond. A – relaxed model. B, C - comparison of experimental and calculated XANES spectra of Xe (B) and Kr (C) in nanodiamond.

Comparison of experimental XANES spectra of Xe implanted into nanodiamonds with sizes 4-5 nm with calculated spectrum shows that the Xe-V model is a plausible possibility (Fig. 8b). A small difference between the position of absorption features in experimental and calculated spectra indicates that in the real sample the distances between Xe and surrounding carbon ions are slightly smaller, than in the model. Interestingly, in bulk diamond only a rather small (~10%) fraction of implanted Xe ions forms stable luminescing complexes (presumably, Xe-V) and annealing to temperature of at least 800 °C is required to anneal radiation damage and drive the implant to stable lattice position.[47,48] However, the heat released in the interaction of the implanted ion with a nanodiamond grain[14-16] might be sufficient to induce lattice reconstruction and formation of the Xe-V. Our attempt to observe room temperature photoluminescence of Xe-related defects using 405 nm excitation was not successful, but it is possible that employed ion fluence was excessive (see dependence of the Xe-defects luminescence from ion fluence in Ref. 52)

The model for isostructural Kr-V defect was constructed, the nearest C atoms are at 2.1 Å. In contrast to the situation with Xe, experimental and calculated XANES spectra of implanted krypton differ markedly (Fig. 8c). Except clearly observable White Line (a sharp feature above the Kr absorption edge), the experimental spectrum is rather featureless. Although some similarity with the calculated spectrum can be recognised (sharp white line, a hump at 14340 eV), the comparison is rather poor. A plausible explanation is that krypton in nanodiamond does not have a single well-defined site, instead several configurations are present; superposition of their contributions leads to smearing of experimental spectrum.

Somewhat surprisingly, the Xe environment in ultrananocrystalline diamond film markedly differs from that in nanodiamond, instead its XANES spectrum resembles that for Xe in graphite, see Figure 9a. From the structural point of view, UNCD consists of elongated diamond crystallites, enveloped into thin (few nm) graphite-like sheath, but volumetrically the diamond phase dominates. Whereas unambiguous explanation is yet lacking it is possible that loss of energy of impinging Xe ions during passage through the graphitic shell prevents implantation into the diamond crystallites and/or damaged region in the diamonds resembles graphitic carbon. Diamond samples retain a smaller amount of Xe in comparison with graphitic C, which serves as indirect support of scenario suggested for the UNCD film. Apparently, higher density and displacement energy of diamond in comparison with graphite both prevents incorporation of ions with small energies and complicates the formation of stable atomic configuration for implanted ones.

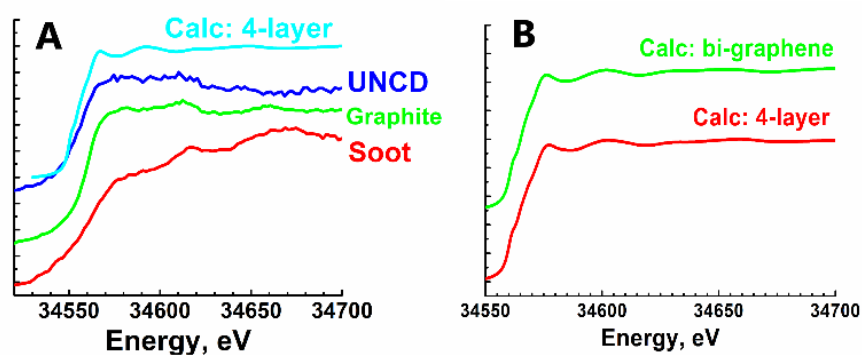

**Fig. 9.** Experimental and calculated XANES spectra of Xe in sp2-C and in UNCD film.

Except H-termination model in Quantum Espresso, no specific surface functional groups on nanodiamond grains were explicitly taken into account in this work. In case of Xe implantation into a grain, the ion slows down with small deviations from initial trajectory.[16] When implanted into central part of a grain, the ion will be surrounded by bulk-like diamond and influence of surface groups will be negligible. We recall here that even in nanodiamonds as small as 1.6 nm and covered with various functional groups, silicon-vacancy defects still serve as efficient photon emitters implying close similarity of their local environment with bulk diamond.[53] Although the Xe-V defect may be formed at any depth in a nanodiamond grain,16 according to quantum chemistry modelling the energy of the Xe-V formation in H-terminated diamond grain is only slightly dependent on the grain size, but moderately decreases towards periphery of the grain.14 This suggests that upon annealing Xe ions may migrate towards grain surface or extended defects. Subsequently, pronounced graphitisation of a nanodiamond surface – here the UNCD may serve as an end-member with fully graphitised surface – may influence stability of the Xe-V defects and significant fraction of Xe may leave the $sp^3$ core. However, close similarity of experimental XANES spectrum with calculated "bulk-like" Xe-V model indicates that in case of nanodiamond grains subjected to standard acidic cleaning the effect of functional groups is minor.

*XANES spectra of Xe in graphene stacks*

Figure 9a shows experimental XANES spectra of Xe implanted in polycrystalline graphite, carbon soot and Ultrananocrystalline diamond (UNCD); for comparison, spectra of Xe ion placed in bi- and 4-layer graphene stack calculated using Artemis software[37] for configurations shown in Fig. 2 are also shown. The calculated spectra of Xe in 2- and 4-layer graphene stacks (Fig. 9b), and of Xe ion in the direct vicinity of a vacancy in one of the graphene sheets are rather similar, which is explained by the rapid decrease of a cross-section of photoelectron-carbon ion interaction with distance, therefore, immediate neighbours dominate the backscattering signal. The experimental spectrum of Xe in graphite and in UNCD film are similar to the calculated ones for graphene stacks, but, as well as the nanodiamond case discussed above, the Xe-C distances in the studied samples are shorter than in the modelled geometry, likely reflecting larger relaxation of real nanocarbons and possible surface reconstruction, not fully encompassed in our calculations.

Carbon soot loses Xe under the beam, which is manifested as a gradual decrease of the absorption, but this process reaches saturation after several tens of minutes under the X-ray beam. Possibly Xe site also evolves during the initial stages of photon beam irradiation. The resulting spectrum of Xe in soot markedly differs from that in graphite. Such discrepancy is not very surprising, since the structure of carbon soot is very complex and although graphene stacks of variable sizes and thickness are present, a significant sample-dependent fraction represents highly disordered carbon. Subsequently, both implanted Xe may occupy a rich variety of sites and a simple model of an ion in between two (or four) graphene sheets fails to explain all the complexity.

**Conclusions**

This work presents results of combined computational and experimental investigation of low energy implantation process and subsequent behaviour of heavy noble gases (Ar, Kr, Xe) in few-layer graphene stacks and nanodiamonds. Behaviour of He and H as well as influence of functional groups on behavior of implanted ions are briefly considered. Good correspondence between calculated and experimental X-ray absorption spectra of implanted Xe is an important support of hypothesis that a Xe-V (Xe-vacancy) complex is the most stable site for Xe ion in diamond lattice; for Xe in nanodiamond this was experimentally shown for the first time. Surprisingly, krypton behaviour is more erratic; possibly several sites may form during the implantation and thus experimental XANES spectra average over several configurations.

Xe implanted into Ultrananocrystalline diamond film is likely expelled from the diamond phase into the graphitic envelopes. Comparison of experimental data and computational results indicate that noble gases introduced in thin graphitic stacks deform neighbouring graphenes, forming bulges. Formation of the bulge appears to be energetically unfavourable for thick stacks due to interlayer interaction, thus explaining why ordered meteoritic graphite is often devoid of the noble gases.[54] The diffusivity of implanted ions in flat and curved graphene stacks markedly depends on their curvature. Whereas incorporation into the flat regions is energetically more favourable, if the ion is trapped in the concave region, significant energy is required to expel it from the trap. The trapping effect is more pronounced when the curved sheets are defective. Thus, localization of noble gases in highly curved domains of thin graphene stacks present in the carbonaceous phase from meteorites may explain their high release temperatures.

## Author Contributions

A.L.T. – Modelling of "static" case; participation in synchrotron experiment; E.N.V. – modelling of implantation processes; K.O.K. – synchrotron experiment; V.L.B. – ion implantation; A.A.S. – concept of the study, ion implantation, synchrotron experiment. All authors participated in data evaluation and manuscript preparation

## Conflicts of interest

There are no conflicts to declare.

## Acknowledgements

Authors thank HZDR and ESRF for the beamtime allocation at BM20 beamline. The quantum chemistry calculations were carried out using high-performance computing resources of the federal collective usage centre Complex for Simulation and Data Processing for Megascience Facilities at NRC "Kurchatov Institute" (http://computing.nrcki.ru/). Simulations of ion implantation processes were performed using equipment of the shared research facilities of HPC computing resources at Lomonosov Moscow State University. We thank Svetlana Trubina from Nikolaev Institute of Inorganic Chemistry SB RAS, Novosibirsk, Russia for providing the grazing-incidence setup for the XANES measurements. A.L.T. and K.O.K. acknowledge support from the Russian Ministry of Science and Education under grant no.075-15-2019-189. Comments of three anonymous reviewers are highly appreciated.